# An Ultra-compact In-line Polarimeter

J. P. Balthasar Mueller,[1] Kristjan Leosson,[2,3] and Federico Capasso[1,*]

[1]Harvard School of Engineering and Applied Science, Cambridge, MA 02138
[2]Innovation Center, Reykjavik, Iceland
[3]Science Institute, University of Iceland, Reykjavik, Iceland
*Corresponding author: capasso@seas.harvard.edu

In-line polarimeters perform non-destructive polarization measurements of optical signals, and play a critical role in monitoring and controlling the polarization environment in for example optical networks. While current in-line polarimeters are constructed with multiple optical components, either fabricated into an optical fiber or using free-space optics, we present here a novel architecture conducive to monolithic on-chip integration. This enables the scalable fabrication of high-performance polarization sensors with exceptional stability, compactness and speed. The method relies on the detection of the highly polarization-dependent scattered field of a sub-wavelength antenna array known as metasurface, and is shown here to provide polarization state measurements matching those of a state-of-the-art commercial polarimeter.

## 1. INTRODUCTION

Polarization characterizes the vectorial nature of electromagnetic (EM) radiation, which represents a fundamental property separate from its frequency and intensity. The measurement of polarization reveals rich information, for example about the structure and composition of materials [1], the handedness of chiral molecules [2], and generally about the nature of scattering, emission, and absorption phenomena [3-6]. Polarization can for example be used to ascertain the texture and orientation of surfaces in remote sensing applications, and help defeat fog, camouflage and image clutter [7]. It also represents an important control parameter in engineering light-matter interactions, including in waveguiding [8-13], nanofabrication [14,15] and biomedicine [16]. Several widely employed characterization techniques rely primarily on polarization measurements, such as ellipsometry and chiral sensing [17,18]. Polarization sensitivity is also used to substantially enhance the functionality of optical and radiofrequency technologies, such as in polarization spectroscopy, microscopy, imaging and radar systems. Unchecked polarization effects can substantially impair network performance in optical telecommunications, which calls for expansive polarization monitoring as the demand for bandwidth rises [19,20]. At the same time, polarization can also be used to improve network bandwidth in polarization diverse systems, and plays a central role in nascent quantum information technology [21-23]. Despite this vast technological potential polarization is nevertheless seldom measured, compared to, for example, intensity or frequency. This may in part be explained by the difficulty associated with capturing its inherently vectorial nature: polarization measurements require several measurements of the same signal, each targeting one of its vector components. This issue is addressed by current polarimeters by dividing the signal up either in space or in time [7,24]. Typically, the signal is either split into several beam-paths (amplitude-division), spread over an array of analyzers (wavefront-division), or measured multiple times with a time-varying analyzer (time-division). Consequently, polarimeters based on conventional discrete optical components quickly grow too large or expensive for many applications, such as the monitoring of thousands of fiber links or in-vivo polarization sensing. Frequently, the need for a birefringent medium to measure polarization helicity can additionally drive up minimum cost and size of polarimeters, and suitable materials are not readily available for all wavelength ranges.

Nanophotonics research has in recent years demonstrated ultra-compact optical components based on planar sub-wavelength structures that are capable of engineering the phase front of reflected and/or refracted light. Such "metasurfaces" can mimic bulk optics, but also achieve new functionality [25]. The latter notably includes the ability to reduce an entire system of bulk optical components to a single, ultra-thin optical element, for example achromats [26] and high-NA aspheres [27]. The flexibility in tuning metasurface functionality through their structure can also bypass the reliance on special material properties, such as birefringence [15,28]. Metasurfaces consequently represent an opportunity for polarimetry to overcome unwieldy and expensive architectures [29-36], but implementations hitherto suffered from parasitic losses, low efficiency or a need for digital data processing. Some designs also partially negate the size-advantage of metasurfaces by requiring detection of light in the far-field. We demonstrate in this article a polarimeter that benefits from the flexibility, simplicity and extreme compactness of a metasurface without introducing such drawbacks. Our design deliberately avoids the use of surface plasmon propagation to minimize optical losses, and relies instead on highly directional scattering to directly measure the state of polarization.

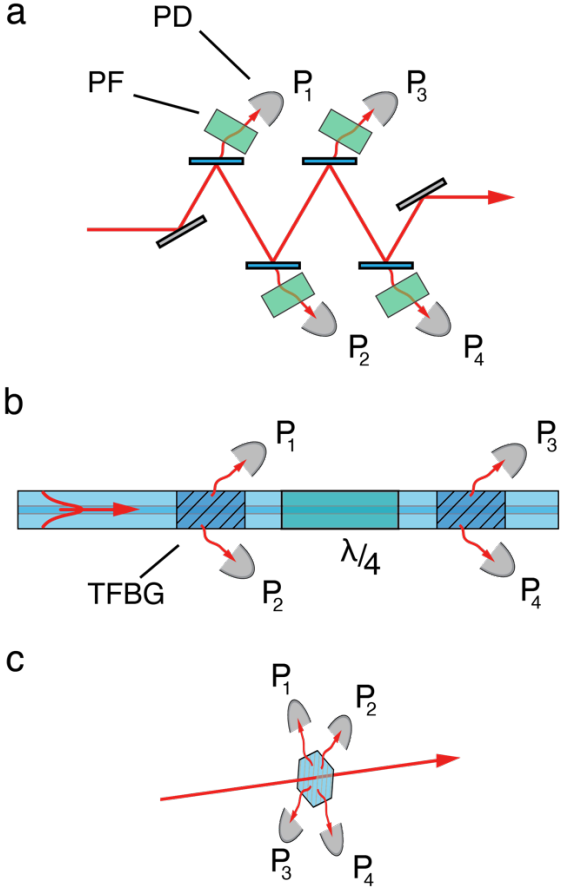

Figure 1: In-line polarimeter architectures. (a) In-line polarimeter architecture based on discrete optical components. The signal (red) is reflected off several semi-transparent mirrors, each time transmitting a fraction of the signal. Different polarization components are measured by applying different polarization filters (PF) to the transmitted intensities and measuring them with photodetectors (PD). (b) Fiber-based in-line polarimeter as used in optical networking. Tilted fiber Bragg gratings (TFBG) act as polarization beam splitters, coupling out beams corresponding to two different polarization components of the signal. Measuring the beams scattered by two TFBGs separated by a fiber-waveplate yields sufficient information for determining the polarization state in the fiber. (c) A subwavelength antenna array can be engineered to generate four (or more) scattered beams, related in intensity to different polarization components of the incident signal. Fast polarization measurements are possible by directly detecting those beams, resulting in an extremely compact architecture with a single, essentially two-dimensional, optical element and four (or more) photodetectors.

Practical polarimeter designs generally fall into one of two categories; either the entire signal is partially filtered, e.g. by rotating polarizers, and the remaining intensity fully converted into photocurrent (destructive measurement), or small fractions of the incoming signal are selectively split off for polarization measurement, leaving the original signal largely unperturbed (non-destructive measurement). Non-destructive, or 'in-line' polarimeters are needed for example for feedback-driven polarization-generation, or when a signal must be monitored live without absorbing it, such as in optical networking. The polarimeter we introduce here performs non-destructive measurements upon transmission through a 2D array of subwavelength antennas, capitalizing on the weak scattering of the antennas to minimize perturbation of the signal.

Conventional in-line polarimeters may use partially transmissive mirrors [37] (Fig 1a), which however do not generally preserve the polarization of the incident beam. These polarimeters measure the components of the polarization vector through amplitude division by sampling the signal multiple times as it propagates through the polarimeter. Each time, a different polarization component is filtered using polarization optics, and measured using photodetectors. Fiber-based in-line polarimeters, which represent the most common form of in-line polarimeters, may use two tilted fiber Bragg gratings separated by a fiber-equivalent of a λ/4 waveplate to split off fractions of the propagating signal (Fig 1b).

## 2. CONCEPT

In contrast to existing architectures, the in-line polarimeter presented here relies only on a single, ultra-thin optical element (Fig 1c). The element, consisting of a 2D metasurface, causes highly polarization-selective directional scattering of the signal as it is transmitted through the polarimeter. The polarization information can then be deduced by directly measuring the intensity of the scattered field at a number of discrete points in space, requiring no polarization filtering. Crucially, the scattered field can be sampled co-planar and very close to the metasurface, which distinguishes it from designs that rely on the far-field detection of diffraction orders. This facilitates dense, planar integration that fully capitalizes on the compactness of the metasurface approach. The design of the antenna arrangement is based on straightforward and intuitive design rules, rather than complex numerical modeling, and can be adapted to a different wavelength range through simple scaling.

The polarization state of electromagnetic waves is historically often described using a 4-element column vector containing the Stokes parameters $\mathbf{S} = (S_0, S_1, S_2, S_3)^T$, where superscript $T$ denotes the matrix transpose [1,19,38,39]. This may be rewritten $S = (I, pIs_1, pIs_2, pIs_3)^T$, where $I$ is the intensity of the wave, $p \in [0,1]$ its degree of polarization (DOP), and $s_1$, $s_2$ and $s_3$ are the components of a 3D unit vector $\mathbf{s} = \hat{\mathbf{s}}_1 s_1 + \hat{\mathbf{s}}_2 s_2 + \hat{\mathbf{s}}_3 s_3$ that characterizes the state of polarization (SOP) of the wave. Through the description as a unit vector, all possible SOPs may be geometrically represented as a point on a unit sphere called Poincaré sphere (Fig 2a). A polarimeter generally functions by performing $n$ independent optical power measurements $P_1, P_2, \ldots P_n$ corresponding to projective measurements of the Stokes vector of the signal $\mathbf{S}$. Assuming linear response, the set of power measurements is related to the Stokes vector of the signal $\mathbf{S}$ through an $n \times 4$ device matrix $\mathbf{M}$ as $\mathbf{P} = \mathbf{MS}$, where $\mathbf{P} = (P_1, P_2, \ldots P_n)^T$. Each row of the matrix $\mathbf{M}$ then corresponds to the Stokes vector of a polarization component that is sampled by the polarimeter. The polarization state is determined from the power measurements by inverting the device matrix, via $\mathbf{S} = \mathbf{M}^{-1}\mathbf{P}$. The accuracy and extent to which polarization can be deduced depends on the properties of the device matrix with respect to inversion, and the left inverse of the device matrix, $\widetilde{\mathbf{M}}^{-1} = (\mathbf{M}^T \mathbf{M})^{-1} \mathbf{M}^T$, can be used instead of $\mathbf{M}^{-1}$ when the dimensions of $\mathbf{M}$ are not $4 \times 4$ [24]. The optimization of the device matrix is a problem that is generally applicable to polarimeter design, not just the architecture discussed here, and is discussed in greater detail in the supplementary information section on device matrix design (Supplement 4). An excellent treatment can be found in particular in Ref. [39]. For the purpose of the current study, we are interested in measuring the SOP of light directly at the end facet of a single mode optical fiber, illustrating the exceptional compactness of our polarimeter design, and its application to the technologically relevant problem of in-line SOP monitoring in fiber-based telecommunications. The antenna array is designed such that it scatters a small part of the normally incident signal directionally in four different directions co-planar with the array, with the intensity scattered in each of the directions proportional to the strength of a different polarization component. The respective measurements $P_1, ..., P_4$ are then made by detecting the power propagating into each direction, and the linear relationship of these values to the incident polarization state $\mathbf{S}$ gives the instrument matrix of the polarimeter $\mathbf{M}$. In order for the inversion of $\mathbf{M}$ to allow for the complete determination of the SOP, the four scattered polarization components need to form a basis for the SOP-space. That is, the set of power measurements must uniquely determine the signal SOP vector. A particular difficulty lies here in making the structure sensitive to polarization helicity (the $\hat{s}_3$ component). This issue was recently addressed in the context of unidirectional surface plasmon-polariton propagation [8,9,13]. We configure our antenna array to have rows of the device matrix $\mathbf{M}$ correspond to four elliptically polarized states with different helicities and azimuths, as shown in Fig 2a. Measurement of these four polarization components unambiguously determines the location of the signal SOP on the Poincaré sphere.

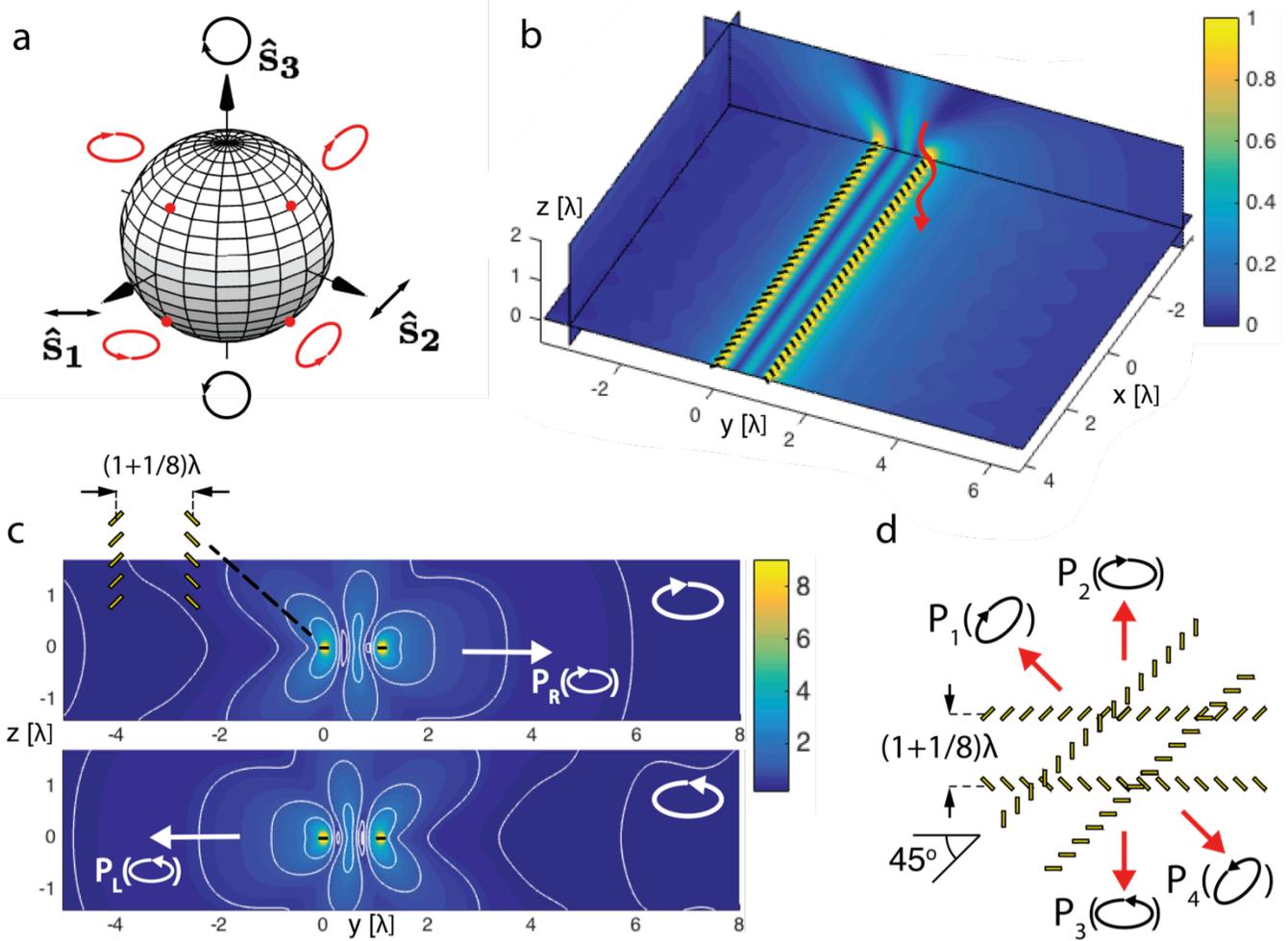

Figure 2: Polarization-selective directional scattering. (a) The Poincaré sphere representation of state of polarization as a point on the unit sphere. The four elliptical polarization components directionally scattered by the antenna array are shown in red. Measuring these components unambiguously defines the state of polarization of a polarized signal. Note that orthogonal polarization states correspond to diametrically opposite points on the sphere. (b) Illuminated at normal incidence, two parallel rows of subwavelength antennas with $\pm 45°$ orientation with respect to the row axis emit a radiation field that is homogenous along the axes of the rows, but polarization-dependent in its radial distribution. The field intensity is here calculated approximating the antennas as perfect non-interacting dipoles, and shown on a colorscale in arbitrary units. (c) When the rows are laterally displaced by $(1+1/8)\lambda$, the row pair directionally scatters elliptically polarized light, with the direction depending on the handedness of the incident polarization. The field intensity is here calculated as in (c) and shown in arbitrary units. (d) Two such pairs of rows superimposed at a $45°$ relative angle result in polarization-selective directional scattering of the desired four elliptical polarization states marked in (a).

Subwavelength rod antennas emit approximately the field of an electric dipole when electromagnetic radiation polarized parallel to their axis is incident upon them. Many such antennas placed in a row with subwavelength spacing will collectively emit a cylindrical wave when radiation is normally incident on the row and polarized parallel to the constituent antennas (see Supplement 1). The waves emitted by two parallel rows of subwavelength antennas with orientation $\pm 45°$ with respect to the row axis will interfere to form a radiation field that is asymmetric in the lateral direction (Fig 2b). The directionality of this radiation field depends on both amplitude and relative phase of the orthogonal linear polarization components that drive each antenna row (see Supplement 2).

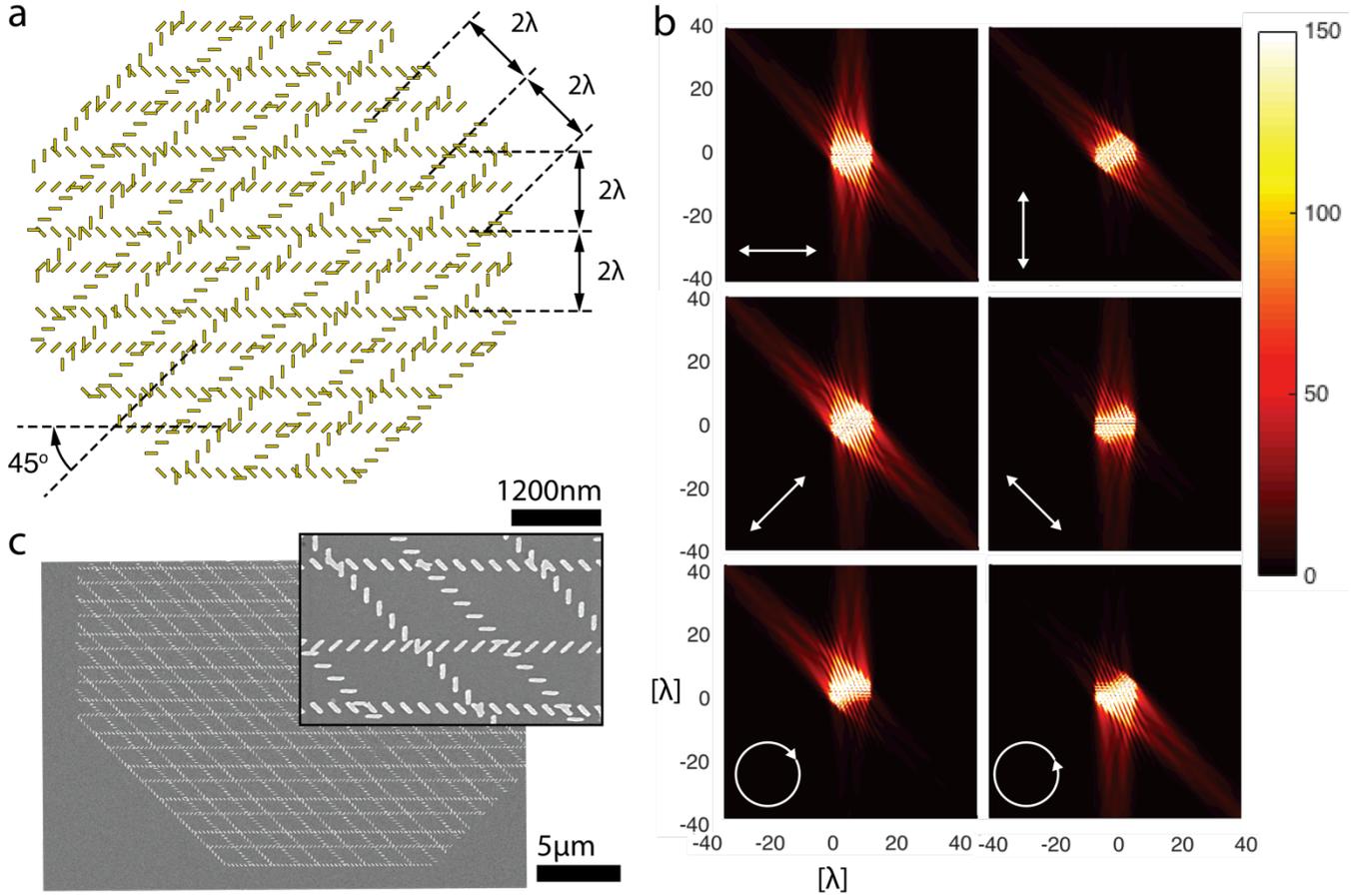

Figure 3: Polarimetric antenna array. (a) The cross-section of the polarimeter is increased by patterning several pairs of antenna rows with a separation of $2\lambda$, resulting in a patch as shown. (b) The calculated in-plane scattered field intensity of a polarimetric antenna array with 5 pairs of rows under illumination with different polarizations in the independent-dipole approximation (in arbitrary units, with the colorscale saturated in the central region). The spatial scale is in units of wavelength. The incident polarization states are shown as white arrows and correspond to the cardinal directions on the Poincaré sphere. Each polarization state results in a unique distribution of intensities over the four beams, implying that their measurement enables the deduction of a signal's state of polarization. (c) Scanning electron micrographs of the fabricated structure, designed for operation at telecommunication wavelengths. Inset: Close-up showing individual antennas of the array.

As special cases, antenna rows spaced by $(m+1/2)(\lambda/2)$, where $m$ is an integer, will directionally scatter circularly polarized light of different handedness in opposite directions [9], while rows spaced by integer multiples of $\lambda/2$ will maximally scatter light that is linearly polarized along the row axis equally in both directions. For other spacings (in the present configuration, we use a row spacing of $(1+1/8)\lambda$), left- and right-handed *elliptically* polarized light is directionally scattered in opposite directions (Fig 2c). Two pairs of rows superimposed at a relative angle of $45°$ may then selectively scatter the required four elliptical polarization states shown in Fig 2A in different spatial directions (Fig 2d), as a rotation of the row of antennas by 45° with respect to the incident radiation corresponds to a 90° rotation of the scattered SOP vectors in the equatorial plane of the Poincaré sphere (see the Supplementary information on device matrix implementation).

The cross-section of the structure may be enlarged by repeating the row pairs every $2\lambda$ in the manner of a grating, resulting in an array of antennas (Fig 3a). This has the side effect of modulating the scattered field of the row pairs into grating orders, which, for normally incident light at $\lambda$, are co-planar and $\approx 60°$ out-of-plane with the polarimeter (see Supplement 3). Figure 3b shows the predicted in-plane scattered field of such an array for different incident polarization states, each resulting in a unique intensity distribution. The polarization response of the actually implemented antenna array will deviate, e.g., due to internal reflections, wavelength dependence, antenna interactions and possible geometric deviations from the intended design. This necessitates a standard calibration experiment in which the polarization-selectivity of the different channels $P_1 - P_4$ in the actual fabricated device across the range of operating wavelengths is determined precisely (see the Supplementary Information).

## 3. EXPERIMENT

For the purpose of our experiment, the four beams scattered by the antenna array are sampled using outcoupling gratings situated in-plane at a distance of 500 μm from the array and picked up in the far field with an imaging detector. This was done primarily to simplify the laboratory setup. In an actual packaged polarimeter device, however, the intensity measurements are more conveniently carried out directly in the configuration depicted in Fig 1c.

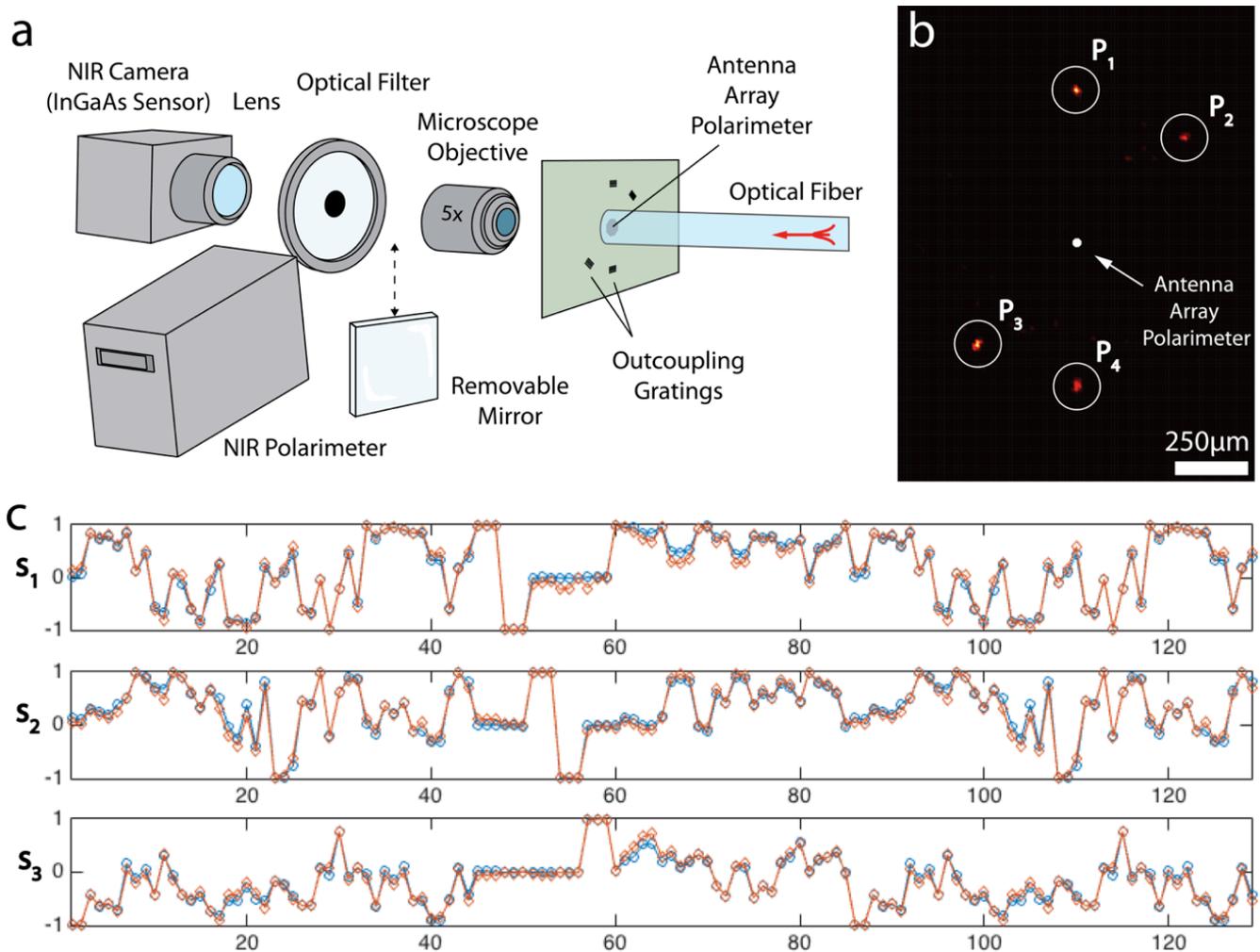

Figure 4: Experimental implementation of the antenna array polarimeter (a) Setup for characterizing the antenna array polarimeter. The array is positioned at the endfacet of an optical fiber, which carries light from a tunable laser source at telecommunication wavelengths. The state of polarization of the light is modified by straining the fiber, and monitored using a commercial NIR polarimeter via a removable mirror. The intensity of the light scattered by the outcoupling grating is measured by imaging them using a NIR camera with an InGaAs sensor, where the light directly transmitted through the metasurface is shadowed with an optical filter. (b) Camera image of the outcoupling gratings, showing polarization-dependent intensities $P_1$, $P_2$, $P_3$ and $P_4$ scattered by the four outcoupling gratings marked with white circles. The antenna array is shown to scale as a white overlay (c) Measurement of the state of polarization $(s_1, s_2, s_3)$ of 129 arbitrarily selected polarizations using the commercial polarimeter (blue) and the metasurface polarimeter (orange).

We fabricate a polarimeter with four output channels and four corresponding outcoupling gratings on top of a double-side polished silicon wafer that was spin-coated with a 12-μm thick layer of benzocyclobutene polymer (BCB, $n \approx 1.535$ at 1550 nm). The antenna array and the outcoupling gratings were patterned using electron beam lithography, followed by electron beam deposition of Ti (1 nm) + Au (21 nm) and liftoff (Fig. 3c). The structure was then covered with a second 12 μm layer of BCB polymer by spin-coating. The resulting individual antennas had a size of approximately 250x50nm. The antenna array was designed for a center operating wavelength of $\lambda = 1550\,\text{nm}/1.535 = 1010\,\text{nm}$.

In the experiment, light from a tunable laser source (Tunics Plus) was guided towards our polarimeter using a single mode optical fiber (SMF28) that was placed in mechanical contact with the BCB surface. The structure was imaged through the silicon wafer using an InGaAs camera, where the light directly transmitted through the polarimeter was blocked in order to avoid saturating the camera detector. The SOP of the incident light could be changed arbitrarily by straining the optical fiber and simultaneously be monitored using a commercially available rotating-waveplate polarimeter (Thorlabs PAX5710IR3-T) using a removable mirror (Fig. 4a). The camera image shows the scattering from the outcoupling gratings as bright spots that change in intensity in response to altering the polarization of the incident light (Fig. 4b) (see Supplementary Fig. S1). We tested the polarimeter at several wavelengths between 1500 nm and 1565 nm, covering the C-band telecommunication wavelengths (1530-1565 nm). Calibration measurements were performed for each wavelength to characterize the precise polarization response of the polarimeter, as described in the supplementary information (see Supplement 5). A number of independent SOP measurements performed using the antenna array polarimeter and the commercial polarimeter were compared for multiple arbitrary polarizations, representative of all possible SOPs on the Poincaré sphere. The results for measurements for incident light at a wavelength of 1550 nm are given in Fig. 4c, with similar plots for wavelengths between 1500 nm and 1565 nm provided in the supplementary information (Supplementary Figs. S2-S4). In all cases, excellent agreement is observed between the SOP measurements derived from the antenna array and from the rotating-waveplate polarimeter. Our design is clearly superior, however, in terms of potential compactness, speed and stability. Fully packaged, the performance of the antenna array polarimeter may well match that of state-of-the-art fiber based in-line polarimeters, where sampling rates and sensitivities are detector-limited, while additionally offering much simpler construction and applicability for free-space radiation as well as other wavelength ranges where detectors are available.

The layout of the antenna array used in the present work was intended primarily for SOP measurements, but also allows for intensity measurements since fully polarized laser light with unchanging DOP was used. Intensity fluctuations manifest themselves as homogenous changes in all channel intensities, and corresponding data is provided in Supplementary Fig. S5. Furthermore, more complex antenna array designs that enable DOP measurements are provided, which rely on six or more outputs (which do not all have to be read out). For the present purpose, the simplest possible, four output design was however adequate.

## 4. CONCLUDING REMARKS

In conclusion, we have demonstrated a fundamentally new architecture for practical, non-destructive polarization measurements based on a single two-dimensional array of rod antennas. The concept relies on the detection of polarization-selective directional scattering. By reducing the polarimeter to a single, ultra-compact optical component and four or more detectors, the proposed architecture can substantially outperform existing polarimeters in terms of size, cost and complexity. The antenna arrays can furthermore be scaled to operate at most technologically relevant wavelengths, enabling polarimetry in wavelength ranges where it was previously very difficult, such as the mid-IR. Through these improvements, our polarimeter design promises to make in-line polarization measurements accessible to a much broader spectrum of applications with portable, mass-produced ultra-compact polarimeter devices.

**Funding**. Air Force Office of Scientific Research (MURI: FA9550-14-1-0389), Iceland Research Fund (152098-051), Thorlabs Inc.

**Acknowledgment**. We thank Alex Cable of Thorlabs Inc. for helpful advice and Michael Juhl for his help with simulations.

See Supplementary Material for supporting content.